\documentclass[12pt]{iopart}
\begin{document}
\begin{flushright}
NIKHEF/99-033 \\
%hep-ph/9912350 
\end{flushright}
\jl{4}

\title[NLO Calculations for Charm Production in DIS]{
NLO Calculations for Charm Production in DIS}

\author{Eric Laenen\dag}

\address{\dag\ NIKHEF Theory Group, Kruislaan 409, 1009 DB,
Amsterdam, The Netherlands}

\begin{abstract}
I present a short overview of the NLO QCD calculations
available for deep-inelastic production of heavy quarks.
\end{abstract}

%\submitted

%\maketitle

\section{Introduction}

Charm quarks produced in deep-inelastic scattering (DIS) have been identified in
sizable numbers now by the H1 \cite{Adloff:1996xq} and ZEUS \cite{Breitweg:1997mj} 
collaborations at HERA, and
considerably more charm (and bottom) data are anticipated.
At the theoretical level the reaction has already been studied extensively.
In the framework where the heavy quark is not treated as a 
parton, leading order (LO) \cite{Witten:1975bh,Gluck:1979aw} and next-to-leading order 
(NLO) \cite{Laenen:1993zk,Riemersma:1995hv} calculations of the inclusive structure functions
exist. Moreover, LO ({\sc AROMA, RAPGAP}) \cite{Ingelman:1997mv,jung:rapgap}  and 
NLO ({\sc HVQDIS}) \cite{Harris:1995tu,Harris:1996hq,Harris:1998zq}
Monte-Carlo programs, allowing a much larger class of observables
to be compared with data, have been constructed in recent years. 
Overall, the NLO QCD description agrees quite well with the HERA data.
Here I shall give a very brief overview of these NLO calculations. 

Charm quarks are produced in DIS via the reaction
\begin{equation}
e^\pm(p_e) + P(p) \rightarrow e^\pm(p_e-q) + 
X[Q,\bar{Q}]\,,\label{one}
\end{equation}
where $P(p)$ is a proton with momentum $p$, $Q$ 
is a heavy quark with momentum $p_1$ ($p_1^2 = m^2$) and
$X$ is any hadronic state allowed, containing the heavy
quark-antiquark pair. Its differential cross section
may be expressed in general as
\begin{eqnarray}
\label{three}
\fl \frac{d^{2+n}\sigma}{dxdQ^2\prod_i dV_i} &=& 
\frac{2\pi\alpha^2}{x\,Q^4}
\Bigg[ ( 1 + (1-y)^2 )
\frac{d^n F_2}{\prod_i dV_i}(x,Q^2,m^2,V_i) 
-y^2 \frac{d^n F_L}{\prod_i dV_i}(x,Q^2,m^2) \Bigg]\,,
\end{eqnarray}
where 
\begin{equation}
Q^2 = -q^2\,,\qquad x = \frac{Q^2}{2p\cdot q} 
\,, \qquad y = \frac{p\cdot q}{p \cdot p_e}\,. \label{four}
\end{equation}
The $V_i$ stand for kinematic variables related
to the heavy quarks. Examples are the transverse momentum 
of the heavy quark, the rapidity difference between the heavy quarks, etc.

\section{Inclusive and single-charm inclusive production}

The least difficult cross section to measure is the
heavy quark inclusive cross section, expressed via 
(\ref{three}) in the 
inclusive structure functions $F_2$ and $F_L$.
These were calculated to NLO in \cite{Laenen:1993zk}. 
The results are parametrized as
\begin{eqnarray}
\fl F_{k}(x,Q^2,m^2) =
 \frac{Q^2 \alpha_s}{4\pi^2 m^2}
\int_x^{z_{\rm max}} \frac{dz}{z}  \Big[ \,e_H^2 f_g(\frac{x}{z},\mu^2)
 c^{(0)}_{k,g} \,\Big] \nonumber \\ 
+\frac{Q^2 \alpha_s^2}{\pi m^2}
\int_x^{z_{\rm max}} \frac{dz}{z}  \Big[ \,e_H^2 f_g(\frac{x}{z},\mu^2)
 (c^{(1)}_{k,g} + \bar c^{(1)}_{k,g} \ln \frac{\mu^2}{m^2}) \nonumber \\  
+\sum_{i=q,\bar q} \Big[ e_H^2\,f_i(\frac{x}{z},\mu^2)
 (c^{(1)}_{k,i} + \bar c^{(1)}_{k,i} \ln \frac{\mu^2}{m^2})  
+ e^2_{L,i}\, f_i(\frac{x}{z},\mu^2) d^{(1)}_{k,i}  \, \Big]  \,\Big] \,, 
\label{eq:1}
\end{eqnarray}
where $k = 2,L$ and the upper boundary on the integration is given by
$z_{\rm max} = Q^2/(Q^2+4m^2)$. 
The parton densities $f_i(x,\mu^2)\,, (i=g,q,\bar q)$ are 
explicitly identified. The scale $\mu$ is the mass factorization and
renormalization scale.
The $\overline{\rm MS}$ coefficient functions 
$c^{(l)}_{k,j}(\eta, \xi)\,,\bar c^{(l)}_{k,j}
(\eta, \xi)\,,
(j=g\,,q\,,\bar q\,;l=0,1)$
and $d^{(l)}_{k,i}(\eta, \xi)$,
$(i=q\,,\bar q\,;l=0,1)$
depend on the scaling variables $\eta =(s-4 m^2)/4 m^2$ 
and $\xi=Q^2/m^2$, with $s$ the square of the c.m. energy of the
virtual photon-parton subprocess.
This implies that in (\ref{eq:1}) $z=Q^2/(Q^2+s)$. 

In (\ref{eq:1}) we 
distinguish the coefficient functions with by their origin. The 
$c^{(l)}_{k,i}(\eta, \xi),\bar c^{(l)}_{k,i}(\eta, \xi)$
originate from those partonic
subprocesses where the virtual photon is coupled to the heavy quark, whereas
the $d^{(l)}_{k,i}(\eta, \xi)$
correspond to the subprocess where the virtual
photon interacts with the light quark.
Thus the former are multiplied by the charge squared
of the heavy quark $e_H^2$, and the latter 
by the charge squared of the light quark $e_L^2$ respectively (both in units of $e$).
Only the terms proportial to $e_H^2$ contain the gluon density.

To obtain numerical results for the inclusive cross section, it is better to use
instead of the original, rather long expressions \cite{Laenen:1993zk},
the much faster parametrized form \cite{Riemersma:1995hv}.
The lowest order term contains only the gluon density. 
Light quark densities only come in at next order, 
contributing only about $5$\%. This is 
the reason $F_2(x,Q^2,m^2)$ is used in global analyses to constrain the gluon density.
Besides the gluon density,
the main source of theoretical uncertainty in 
$F_2(x,Q^2,m^2)$ is the value of the charm quark (pole) mass,
rather than the scale $\mu$.

The calculation that lead to (\ref{eq:1}) also yielded
the single heavy quark differential cross section, with 
$V_1=p_T^Q$ and $V_2=y^Q$ \cite{Laenen:1993xs}. 
These distributions are best generated using the HVQDIS 
program, described in the next section.

NNLO estimates based on soft gluon resummation 
are given in Ref.~\cite{Laenen:1998kp}, for inclusive
and single heavy quark inclusive cross section.

\section{Fully differential charm production}

A NLO calculation also exists for the fully 
differential cross section in (\ref{three})
\cite{Smith:1996ts}. Maintaining full differentiality 
required a complete recalculation of the matrix
elements, carefully eliminating intermediate
divergences via the so-called subtraction method. 
The results are encoded in the program {\sc HVQDIS} 
\cite{Harris:1996hq} \footnote{See \cite{Harris:1998xc}
for a description of the most recent changes.}.
The program can compute, to NLO, experimentally
visible cross sections, which are in principle better for 
comparison with theory than fully inclusive ones. 
It returns parton kinematic configurations and their corresponding 
weights, accurate to ${\cal O}(\alpha\alpha_s^2)$.
The user is free to histogram any set of 
infrared-safe observables and apply 
cuts, all in a single histogramming subroutine.

Additionally, one may study heavy hadrons using the  
Peterson {\em et al}.\ fragmentation model.
Detailed physics results from this program are given in 
\cite{Harris:1998zq}. 

HVQDIS has been used extensively in
experimental analyses. As is shown elsewhere \cite{Sutton:dur99}
in these proceedings, it reproduces the data very well indeed, 
except for $D^*$'s at low $p_T$ and large pseudorapidity,
where there are more events than HVQDIS would predict.
This is possibly due to remnant beam drag
effects distorting the pseudorapidity spectrum
to larger values. For the case of charm photoproduction
this was investigated in \cite{Norrbin:1999mz}.

A more extensive overview of the NLO calculations
and the phenomenology of DIS charm production
can be found in \cite{Harris:1999it}.

\end{document}